\input amstex
\documentstyle{amsppt}
\loadbold
\magnification=\magstep1
{\catcode`\@=11\gdef\logo@{}}
\pagewidth{16 true cm}
\pageheight{24.2 true cm}
\voffset 0.5 cm
\document
\parindent = 0.5cm
\parskip = 1.5mm
\define\aaa{1.0cm}
\define\bbb{0.8cm}

\define\pa{\partial}

\define\pd#1#2{\frac{\pa#1}{\pa#2}}
\define\po#1{\frac{\pa}{\pa#1}}

\pageno=1
\NoRunningHeads
\TagsOnRight

\vskip 2cm
\centerline{\bf AN AAD MODEL OF POINT PARTICLE AND THE PAULI EQUATION}
\vskip 1 cm
\centerline{J. Weiss
\footnote{email: weiss\@cvt.stuba.sk}}
\vskip 2 cm
\centerline{\sl Department of Physics, Slovak Technical University}
\centerline{\sl N\'am. Slobody 17,  \, 812 31 Bratislava, Slovakia}
\vskip 3 cm
\indent
    The classical relativistic linear AAD interaction, introduced by the author,
leads in the case of weak coupling to a pointlike particle capable to be
submitted to quantization via Feynman's path integrals along the line
adequate to the requirements of the Pauli equation. In the discussed
nonrelativistic case of the model the concept of spin is considered within
early Feynman's ideas.
\vskip 4 cm
\leftline{PACS: 03.65Pm, 11.25.-w, 12.15.-y, 11.15.Ex}

\newpage
\vskip \aaa
\centerline{\bf 1. Introduction}
\vskip \bbb
It is known that Feynmanian quantization is suitable for systems having a
classical analogue. If such an analogue does not exist, e.g. as is the case
for particles with spin, one encounters difficulties. In this context we want
in the present work to illustrate the following facts: 1) the problem of spin
could be attempted by renovating Feynman's spin idea [1], according to
which a unit vector represents the spin germ cell in the nonrelativistic case;
2) this picture on spin could have the cause in Wigner's suggestion of the
classical free point particle with internal degrees of freedom [2]; 3) such
a point of view may have its deeper origin in the weak coupling model of linear
interaction [3],[4],[5], in which the internal variables, products of AAD
theory, adopt the status of canonically conjugate variables; 4) the model
of linear relativistic interaction, even if based on the AAD theory, is able
to yield its quantum versions on the levels both of Dirac [5] and Pauli
equations. The model of above type, verified in [5] with the assumptions
1)-3), suggests that the relativistic variant of Feynman quantum formalism,
leads to the Feynman - Gell-Mann equation [6], equivalent to the
Dirac equation.

The concept of particle, used in [5] and established on the unit vector $\vec
n$, which is associated with the particle internal variables, is the key tool
for the Feynman nonrelativistic way of quantization proposed in the
present article. The model permits to evaluate the changes of this
vector, involved in the given amplitude, as those potentially realizable in
the classical picture. As a consequence, the standard action of the particle
in the electromagnetic field can be expressed in the form of a chain of
$\delta$ functions. Using such the time evolution of $\vec n$, we can
determine the propagator $K_{S}$ for the Pauli equation.

In this work it is also indicated that some reformulation of theory for
particles with the spin 1/2 within the Feynman path integrals, combined
with progressive tools of AAD theory, could shed a new light on the old
question about the structure of particles. An extended formulation of
old theory perhaps does not exclude yet that one can hint a new way of
exploring problems as to the particle spectrum, the difference between
generations of elementary particles or at least why they have no spin
excitated states.

In this paper we describe first briefly (Sect.2) the above weak coupling
AAD model. In the context with it we utilize the postulated property of
canonicity of internal variables and using their constraints we link them
with the Feynman idea on the character of spin, in the strict
sense of Feynman's conception of a classical spin germ cell (Sect.3)
We propose then the quantization continual-integral procedure of the model,
starting  from the standard form of the Pauli equation, adequate to it
(Sect.4).
\vskip \aaa
\centerline{\bf 2. Major properties of the model}
\vskip \bbb
The linear AAD interaction has its beginning in the papers [3],[4]. It was
developed into the group form that underlies the generator procedure of
constrained Hamilton dynamics [7] and [8]. It was shown [8] that the
realization of the Poincar\'e group for the testing particle is compatible
with the Lie algebra of the de Sitter group SO(2,1), proper to rotator models.
The causality of this kind of interaction together with that
of Wheeler-Feynman AAD electrodynamic was studied in [9] (see also [10]).

The scale of problems concerning the complete explication of linear AAD
interaction in both the classical and quantum regions is clearly too wide.
It can be shown, between others, e.g. that this type of interaction
has necessary features, demanded by the so-called Feynman argument [11],
according to which there exist theories which lead directly from
their equations of motion and quantum conditions to the electrodynamic-like
formula for the Lorentz force and two of Maxwell equations.

An attempt to find a picture on the nature of this interaction for weak
coupling - the only region, where one traditionally can understand how
quantum field theories behave - was made in [4]. It was explored that
in the limit of its weak potentials, i. e. if
$$m\gg g^{2}A^{2} ;\quad m^{2}\gg (\vec \pi - g\vec A)^{2} , \eqno(2.1)$$
where $m$ is the mass of the particle, $\vec \pi$ its canonical momentum
and $A^{\mu}$ the linear fourpotential defined as
$$A^{\mu} \equiv (\phi,\vec A) = g(\xi^{\mu} - \eta^{\mu}) , \eqno(2.2)$$
with the coupling constant $g$, the angular momentum $M^{\mu \nu}$ generator
of the object has the form
$$ M^{\mu \nu} = x^{\mu}p^{\nu} - x^{\nu}p^{\mu} + b(\xi^{\mu}\eta^{\nu} -
\xi^{\nu}\eta^{\mu}). \eqno(2.3)$$ In (2.3) the quantities $x$ and $p$ are
the conventional canonical variables and $\xi$ and $b\eta$ ($b$ being a
constant, required to hold the canonicity) are postulated to be
the corresponding internal variables of the particle that is now regarded
(\`a la Wigner; it has namely Wigner's proportion) as a free particle.
Strictly speaking it should be named the canonical weak potential
particle, or briefly only the WP particle.
The Hamilton canonical formalism yields for both the pairs of the canonically
conjugate variables the appropriate equations of motion [4], the internal
variables being subjected to the constraints
$$\xi^{2} = 0 ;\quad \eta^{2} = 0 . \eqno(2.4)$$
Here we are obliged to respect both the validity of (2.4) and the canonicity
of variables and thus to calculate with the Dirac brackets. These brackets
amount $$\{x^{\mu},p^{\nu}\}^{*} = g^{\mu \nu} ; \quad
\{\eta'_{k},\xi_{i}\}^{*} = \delta_{ik} ; \quad \{\xi_{k}, \xi_{i}\}^{*} =
\{\eta'_{k},\eta'_{i}\}^{*} = 0 , \eqno(2.5)$$ if $\vec \eta' =
\vec \eta - {\eta^{0}\over \xi^{0}}\vec \xi$.
As it was shown [4], the generators $\pi^{\mu}$ and $M^{\mu \nu}$ of the
linear interaction are the products of the weak potential (WP) limit. We note
that in the relation (2.3) the mechanical momentum enters in the role
of the canonical one, since in the WP limit $\pi^{\mu}=gA^{\mu}$ and hence
$p^{\mu}=\pi^{\mu}$. Thus the generators $p^{\mu}$ and $M^{\mu \nu}$ obey
(2.5) or its early variant
$$\{\xi^{\mu},\eta^{\nu}\}^{*}=b^{-1}\Bigl (g^{\mu \nu} - (\xi.\eta)^{-1}
\xi^{\nu}\eta^{\mu}\Bigr ),$$ being next $\{\xi^{\mu},\xi^{\nu}\}^{*}=
\{\eta^{\mu},\eta^{\nu}\}^{*}=0.$ This is a notable sign of the considered
model and it can lead to serious physical implications about the character
of AAD interaction at all. It is apparent at first sight that the variables $\xi$
and $\eta$, introduced in [3], evoke to ascribe to them a statical
(position) - dynamical (momentum) meaning as to the internal variables
characterizing the profile of the WP particle. Actually, they express
the intrinsic fact that the dependence on the position of the particle
is merely implicitly hidden in them. Next, they have (the second one
multiplied by the convenient constant) the feature of canonicity a this
property is linked to constraints, just those on the world cone, which
become today attractive from the point of view of quantization.
\vskip \aaa
\centerline{\bf 3. Starting point of quantization: the classical germ
cell of spin}
\vskip \bbb
   Let us introduce into dynamics of the examined particle the unit vector,
which is constructed from the one of internal components of variables $\xi$
or $\eta$. This particle is characterized by the fixed mass $m$, momentum $p$
and position $x$. Beside these variables we shall describe particle states by
the unit vector $\vec n=\vec \xi/\xi$, where $\xi$ is the magnitude of the
vector $\vec \xi$. In the relativistic theory the introduction of the fourvectors
$\xi$ and $\eta$ and with them also $\vec n$ is demanded, if we insist on
the validity of condition (2.4) and require the realization of PG
of the WP particle to be transitive even for $p^{2}\geq 0$. Wigner
secures the transitivity of the generator (2.3) of PG only by
choosing the constant $b$ negative for the representations with $p^2=0$
and equal to zero for $p^2<0$.

In the nonrelativistic theory
the physical interpretation of $\vec n$ can be argued by the Feynman idea
[1], expressed in connection with the classical limit of the Dirac equation.
Feynman suggests the need of existence of a nonzero vector given by the ratio
of spin and magnetic momentum of the particle. In the classical limit this
vector is not cancelled and fulfils the equation characterizing the spin
precession in an external electromagnetic field. Since in the classical limit
both the quantities are equal to zero, the vector expressing the above ratio
decouples from the orbital degrees of freedom and becomes an "isolated"
vector without the proper physical interpretation. It represents, however,
a classical "germ cell" of the spin and activates itself in the state of
transition to quantum theory. $\xi$ and $\eta$ are thus in a sense the
"dumb" variables, which begin to communicate only in the quantum theory.
Hence there is nothing against common sense, if one admits the existence of
a unit vector as an element of description of the point particle states.
And so we associate its existence directly with the internal variables,
to the best of our belief that this conception can have the serious
physical ground, linking the spin with the weak linear interaction.

Even if one cannot yet go into whys and wherefores thorough,
nevertheless we confine ourselves to the nonrelativistic Pauli equation,
which will reflect the Feynman idea on spin as a unit vector and our
assumption that this vector is a classical relic of the internal variables
of the WP particle, discussed in Sect.2. It is obvious that the full relevance
of our suggestion may be revealed in the relativistic form of this equation.
Results of this kind of computations are given in [5].

As is known, quantization via Feynman's integrals is based on the definition
of the action, which determines the phase of the amplitude, provided that
this amplitude belongs to the appropriate path. The continual integrals
for the propagator of the nonrelativistic Pauli equations were studied by
Schulman [12]. In the present case, oppositely to the relativistic one [5],
one can find no sufficiently simple
action which would be able to generate the equation of motion for $\vec
n$. Let us suppose thus that $\vec n$ obeys the equation
$${d\vec n\over dt} = {e\over mc}\vec n \times \vec H , \eqno(3.1)$$
where $e$ is the charge of the particle and $\vec H$ the strenght of external
magnetic field. We suppose here that the particle owns the electric charge,
of course. Eq.(3.1) together with the equation of motion for $\vec x$
$$m{d^2\vec x\over dt^2} = e(\vec E + {\vec v\over c}\times \vec H)
\eqno(3.2) $$ represent the starting equations for Feynmanian quantization.
The form of (3.2) indicates that in the given case we need define no
action for quantization. The time evolution of $\vec n$ is namely governed
by the same classical law as in quantum theory. There is no external
violation of evolution for $\vec n$ and we are able to determine this
vector precisely at any time. This is the opposite situation to the case of
evolution of $\vec x$, where it holds
$${d\vec x\over dt} = m^{-1}(\vec p - {e\over c}\vec {\tilde A}),
\eqno(3.3)$$
where $\vec {\tilde A}$ is the vectorpotential of the external field and
${\vec p}$ cannot be given accurately at the arbitrary time together with
$\vec x .$

This consideration allows to make the conclusion that the amplitude
belonging to the appropriate path is nonzero only if $\vec n$ is changed
along it in the classical way. It may be written down symbolically as follows
$$e^{{i\over \hbar}S}\delta \Bigl (d\vec n - {e\over mc}\vec n \times
\vec Hdt\Bigr ) , \eqno(3.4)$$ where the Dirac delta function assures the classical
equation of motion to be satistied and next
$$S = \int_{t_{1}}^{t_{2}}\Bigl ({m\over 2}{\vec v}^{2} - e\tilde {A^{0}} +
e{\vec v\over c}.\vec {\tilde A}\Bigr )dt \eqno(3.5)$$
is the standard action of the point particle in the electromagnetic field.

Let us explain how to understand the $\delta$ function in (3.4). For this
purpose let us divide the time interval $t_{n}-t_{1}$ into $N-1$ parts of
the lenght $dt$ and denote $\vec n$ at the time $t_{i}$ as $\vec n_{i}$. The
equation of motion for $\vec n$ can be then written down as
$$\vec n_{i+1} = \vec n_{i} - {e\over mc}\vec H_{i}\times \vec n_{i}dt ,
\eqno(3.6)$$ or $$\vec n_{i}'.\vec n_{i+1} = 1 , \eqno(3.7)$$ where
$$\vec n_{i}' = \vec n_{i} - {e\over mc}\vec H_{i}\times \vec n_{i}dt .$$
We see that $\vec n_{i}$ is the vector $\vec n_{i}$ rotated on $\delta
\vec \varphi$ $$\vec n_{i}' = \vec n_{i} - \delta\vec \varphi\times \vec
n_{i} , \eqno(3.8)$$ where $\delta \vec \varphi=(e/mc)\vec Hdt .$
Eq.(3.7) is equivalent to the original vector equation, because due to
the validity $\vec n_{i}^{'2}=1$ and $\vec n_{i+1}^{2}=1$, one deduces
$\vec n_{i}'=\vec n_{i+1}$. The function $\delta$ in (3.4) can be thus
expressed in the chain form:
$$\delta \Bigl (d\vec n - {e\over mc}\vec n\times Hdt\Bigr ) =
\delta (1 - \vec n_{1}'.\vec n_{2})\delta (1 - \vec n_{2}'.\vec n_{3})...
\delta (1 - \vec n_{N-1}'.\vec n_{N}) . \eqno(3.9)$$
The chain of $\delta$ functions, given in (3.9), represents the time
evolution of $\vec n$ as a progressive succession of infinitesimal
rotations of $\vec n$.
\vskip \aaa
\centerline{\bf 4.  Propagator}
\vskip \bbb
   We are interested in the propagator $K$ in the equation
$$\Bigl (i\hbar{\partial \over \partial t} -\tilde H\Bigr )K(\vec x_{2},\vec x_{1},
t_{2},t_{1}) = i\hbar \delta (x_{2} - x_{1})\delta (t_{2} - t_{1}) ,
\eqno(4.1)
$$ corresponding to the Pauli nonrelativistic equation
$$i\hbar {\partial \psi\over \partial t} = \Bigr [{1\over 2m}(\vec p - ec
^{-1})\vec {\tilde A})^{2} + e\tilde A_{0} - {e\hbar \over 2mc}\vec
\sigma.\vec H\Bigr ]\psi . \eqno(4.2)$$
In (4.1) and (4.2) $\tilde H$ and $\vec \sigma$ represent the Hamiltonian and Pauli
matrices, respectively. According to our conception of the internal vector
$\vec n$ the scalar propagator $K_{S}$ will represent the total amplitude
for the transition $\vec x_{1},\vec n_{1},t_{1} \rightarrow \vec x_{N},
\vec n_{N},t_{N}$ and may be obtained by the summation through the particular
paths and time evolutions of $\vec n$
$$K_{S} = \int e^{{i\over \hbar}S_{N1}}\delta (d\vec n- {e\over mc}\vec n \times
\vec H dt) \Cal D  x (t) \Cal D  \vec n (t) ,\eqno(4.3)$$
where as is the rule
$$\Cal D x(t) = d^3x_{2}d^3x_{3}....d^3x_{N-1}(2\pi i\hbar m^{-1} dt) ,$$
$$\Cal D \vec n(t) = {d\Omega_{2}\over 2\pi}{d\Omega_{3}\over 3\pi}....
{d\Omega_{N-1}\over 2\pi}$$ and $d\Omega_{i}$ is the solid angle appropriate
to the unit vector $\vec n_{i}$. As emphasized above, due to the presence of
$\delta$ function in the integration over $\Cal D \vec n(t)$ only
classical histories of $\vec n$ give nonzero contributions. From the amplitude
$K_{S}$ we obtain the final spinor amplitude $K$, which obeys Eq.(4.1). We
must, however, introduce the spinor $\zeta$ as the solution of the equation
$$\vec n.\vec \sigma \zeta = \zeta \eqno(4.4)$$
with $$\zeta = \left(\matrix e^{-i\varphi /2}\cos {\vartheta \over 2} \\
e^{i\varphi /2}\sin{\vartheta \over2} \endmatrix\right ),   $$
where $\vartheta$ and $\varphi$ are the spherical angles of $\vec n$. Then
$$K(\vec x_{N},\vec x_{1},t_{N},t_{1}) = \int K_{S}(\vec n_{N},\vec x_{1},
\vec n_{N},\vec n_{1},t_{N},t_{1})\zeta_{N}\zeta_{1}^{+}{d\Omega_{1}\over
2\pi}{d\Omega_{N}\over 2\pi} , \eqno(4.5)$$ where $\zeta_{1}$ and $\zeta_
{N}$ are the spinors appropriate to $\vec n_{1}$ and $\vec n_{N}$,
respectively.

One can easily show that there is a relation between $\zeta(\vec n_{N-1})$
and $\zeta(\vec n_{N-1}')$. The spinor $\zeta$ is simply transformed as
$$\zeta' = \zeta + {i\over 2}\delta \vec \varphi.\vec \sigma \zeta ,
\eqno(4.6)$$ if the transformation of $\vec n$ is
$$\vec n' = \vec n - \delta \vec \varphi\times \vec n , \eqno(4.7)$$
being $\vec n'.\vec \sigma \zeta'=\zeta'$. The integral over $d\Omega_{N}$
then gives
$$\int \zeta_{N} \delta(1-\vec n_{N-1}'.\vec n_{N}){d\Omega_{N} \over
2\pi} = (1 + {i\over 2}\delta\vec \varphi.\vec \sigma)\zeta_{N-1} ,
\eqno(4.8)$$ and the next one
$$\int \zeta_{N}\delta(1-\vec n_{N-1}'.\vec n_{N})\delta(1-\vec n_{N-2}'.
\vec n_{N-1}){d\Omega_{N}\over 2\pi}
{d\Omega_{N-1}\over 2\pi} = \eqno(4.9)$$
$$=(1 + {i\over 2}\delta\vec \varphi_{N-1}.\vec
\sigma)(1 + {i\over 2}\delta \vec \varphi_{N-2}.\vec \sigma)\zeta_{N-2} .
$$
We see that in this way all of next integrals may be calculated without
the only one, over $d\Omega_{1}$. Thus the full sequence, scanned
by means of formula (4.9), will finish with the expression
$(1+{i\over 2}\delta\vec \varphi_{1}.\vec \sigma)\zeta_{1}$. One can
easily shown that the integral over $d\Omega_{1}$
represents the unit matrix.

Summarizing the mentioned considerations, we finally find that
$$\int {d\Omega_{1}\over 2\pi}\int {d\Omega_{N}\over 2\pi}\int \delta
\Bigl ({d\vec n\over dt}-{e\over mc}\vec n\times \vec H\Bigl )\zeta_{N}
\zeta_{1}^{+}\Cal D \vec n(t) = \eqno(4.10)$$
$$=(1+{i\over 2}\delta \vec \varphi_{N-1}.\vec \sigma)(1+{i\over 2}\delta
\vec \varphi_{N-2}.\vec \sigma)...(1+{i\over 2}\delta \vec \varphi_{1}.
\vec \sigma)=Te^{{i\over 2mc}\int _{t_{1}}^{t_{N}}\vec H.\vec \sigma dt} ,
$$ where T is the symbol of chronological order. For the total amplitude
$K$ we obtain
$$K(\vec x_{N},\vec x_{1},t_{N},t_{1})=\int e^{{i\over \hbar}S}Te^{{ie\over
mc}\int_{t_{1}}^{t_{2}} \vec H.\vec \sigma dt} \Cal D \vec x(t) =
\eqno(4.11)$$
$$=T\int e^{{i\over \hbar}S+i{e\over mc}\int_{t_{1}}^{t_{2}} \vec H.\vec
\sigma dt}\Cal D \vec x(t) .$$
It is evident that this amplitude obeys the Pauli equation (4.2). Let us
add that even the scalar amplitude $K_{S}$ can be considered in the role of
the propagator, if one admits the expression $\zeta^{+}\psi$ to be the
wave function with $\psi$ understood as the standard spinor amplitude of
initial state $$\psi = \left(\matrix \psi_{i}(\vec x,t) \\
\psi_{f}(\vec x,t) \endmatrix\right).$$
In this case the amplitude of transition from the state $\psi_{i}$
to the state $\psi_{f}$ has the form
$$A_{fi} = \int K_{S}(\vec x_{N},\vec x_{1},\vec n_{N},\vec n_{1},t_{N},t_{1})
(\psi_{f}^{+}\zeta_{N})(\zeta_{1}^{+}\psi_{i}){d\Omega_{1}\over 2\pi}
{d\Omega_{N}\over 2\pi}d^{3}x_{1}d^{3}x_{N} . \eqno(4.12)$$

It turns out that the quantization procedure, examined in Sect.4, is possible
to be extended to particles with higher spins. For instance, for complete spin
$s$ it is sufficient to replace the spinor $\zeta$ by the $2s+1$ component
quantity
$$Y_{s}(\vec n) = \left(\matrix
Y_{s1}(\vec n) \\
Y_{s1-1}(\vec n) \\
. \\
. \\
Y_{sm}(\vec n) \\
. \\
. \\
Y_{s-m}(\vec n) \endmatrix \right ),
 \eqno(4.13)$$
where $Y_{sm}(\vec n)$ is the spherical function. Under the infinitesimal
rotations $\vec n$ this function is transformed as follows
$$Y_{s}' = (1 + i\delta \vec \varphi.\vec s) , \eqno(4.14)$$
where $\vec s$ is the spin operator. Consequently, the integrals over
$d\Omega_{\vec n_{i}}$ allow to be calculated as simply as in the case
of spin 1/2. The final integral $\int Y_{s}Y_{s}^{+}d\Omega$ leads also to
the unit matrix.
\vskip \aaa
\centerline{\bf 5. Conclusion}
\vskip \bbb
   As emphasized in Sec.2, we explore in this paper the AAD model of
particle, elaborated from the linear AAD interaction [3] using its weak
potential [4] (denoted workingly by the WP symbol). The WP particle
exhibits features close to Wigner's model of elementary particle [2],[14].
The formal unifying both the models demanded merely to codify the WP
variables $\xi$ and $\eta$ as a pair of canonical variables and give
them the physical meaning of a bridge from the internal variables to spin.

The point character of a particle is more adequate to the early physical
comprehension of elementariness, as accented by F. Rohrlich [13]. However,
one can ask him a question, is there the possibility to construct the model of a
point particle, states of which allow to ignore the Wigner condition imposed
on the irreducibility of representations of the Poincar\'e group [2],[14]
for an elementary particle? Such a possibility seems to be real in AAD theory.
In the Wheeler-Feynman theory the particle pointness is regarded
as the requirement for the particle to be fully described by giving its
complete worldline. Within this theory one can find cases hinting
that the point particle may be correctly described without insisting at
the irreducibility. Hence two above concepts - the pointness and
elementariness - do not appear to be quite equivalent, the former being
of the wider content.

In the AAD approach, underlain by the extended Dirac generator procedure for
constrained Hamilton dynamics [7],[15] and [8]
the generators of motion together with constraints, which are associated
with the model, determine the equations of motion and by help of them
permit to monitore the pointness. In such a way, using the AAD formalism, one
obtains yet a set of other advantages for description of particles. Due to them
the AAD theory of linear interaction [3] offers the model of the WP particle
[4] precisely \`a la Wigner's pattern [2] of free particle with the internal
variables $\xi$ and $b\eta$, obeying the light cone constraints.
In contradiction with Wigner's conception of elementariness, which
imposes some limits on $p^{2}$, the idea of pointness, based on AAD
constraint dynamics with the same constraints, contains no such restrictions.
This model allows, as was seen, to renovate Feynman's idea on spin [1], which
has its nonrelativistic core in the form of a unit vector, composed of
"dumb" internal dynamical variables of our model and ariving on the physics
scene till in the quantum theory. Thanks to this vector, it was possible to
obtain the propagator corresponding to the Pauli equation in Feynman's way of
quantization.

It turns out that there exists a possibility how to associate the model
of Dirac particle, discussed in [5], with the model of top as a string at
low energies [16] (we are indebted to its
author M. Petr\'a\v s - now already died; he will be missed
but not forgotten by our physics community - for useful suggestions) at a
convenient mediatory role in establishing the link between strings and
the spin 1/2 particles.

\vskip \aaa
          \centerline{\bf References}
\vskip \bbb
\noindent
[1] R.P. Feynman, Rev. Mod. Phys. {\bf 20}, 367 (1948). \newline
\noindent
[2] E.P. Wigner, Ann. Math. {\bf 40}, 149 (1939). \newline \noindent
[3] J. Weiss, J. Math. Phys. {\bf 27}, 1015 (1986); {\bf 27},
1023, (1986). \newline \noindent
[4] J. Weiss, Czech. J. Phys. {\bf B39}, 1239 (1989). \newline \noindent
[5] J. Weiss, see the previous paper. \newline \noindent
[6] R.P. Feynman and M. Gell-Mann, Phys. Rev. {\bf 109},
193 (1958). \newline \noindent
[7] E.C.G. Sudarshan, N. Mukunda and J.N. Goldberg, Phys. Rev.
{\bf D23}, 2218 (1981). \newline \noindent
[8] J. Weiss, Acta Phys. Pol. {\bf B19}, 977 (1988). \newline \noindent
[9] R.A. Moore, D.W.Qi and T.C. Scott, Can. J. Phys. {\bf 70},
772 (1992). \newline \noindent
[10] R.A. Moore and T.C.Scott, Phys. Rev. {\bf A52}, 4371 (1995).
\newline \noindent
[11] F.J. Dyson, Am. J. Phys. {\bf 58}, 209 (1990). \newline \noindent
[12] L.S. Schulman, Phys. Rev. {\bf 176}, 1558 (1968). \newline
\noindent
[13] F. Rohrlich, Am. J. Phys. {\bf 65}, 1051 (1997). \newline \noindent
[14] V. Bargman and E.P. Wigner, Proc. Natl. Acad. Sci. U.S.
{\bf 34}, 211 (1948). \newline \noindent
[15] P.A.M. Dirac, Can. J. Math. {\bf 2}, 129 (1950). \newline \noindent
[16] M. Petr\'a\v s, preprint, lanl.com. hep-th/9909036, 7 Sep 1999.
\newline
\noindent

\end{document}